# PEO/CHCl$_3$: Crystallinity of the polymer and vapor pressure of the solvent
## - Equilibrium and non-equilibrium phenomena -

Albina KHassanova[a)] and Bernhard A. Wolf[b)]

*Institut für Physikalische Chemie der Johannes Gutenberg-Universität Mainz and Materialwissenschaftliches Forschungszentrum der Universität Mainz, Welder-Weg 13, D-55099 Mainz, Germany*

Vapor pressures were measured for the system CHCl$_3$/PEO 1000 (PEO stands for polyethylene oxide and 1000 for $M_w$ in kg/mol) at 25 °C as a function of the weight fraction $w$ of the polymer by means of a combination of head space sampling and gas chromatography. The establishment of thermodynamic equilibria was assisted by employing thin polymer films. The degrees of crystallinity $a$ of the pure PEO and of the solid polymer contained in the mixtures were determined via DSC. An analogous degree of polymer insolubility $b$ was calculated from the vapor pressures measured in this composition range. The experiments demonstrate that both quantities and their concentration dependence are markedly affected by the particular mode of film preparation. These non-equilibrium phenomena are discussed in terms of frozen local and temporal equilibria, where differences between $a$ and $b$ are attributed to the occlusion of amorphous material within crystalline domains. Equilibrium information was obtained from two sources, namely from the vapor pressures in the absence of crystalline material (gas/liquid) and from the saturation concentration PEO (liquid/solid). The thermodynamic consistency of these data is demonstrated using a new approach that enables the modeling of composition dependent interaction parameters by means of two adjustable parameters only.

## I. INTRODUCTION

Polymer solutions of the present type are particularly interesting because of the possibility to obtain thermodynamic information on the interaction between the components from two sources: The chemical potential of the solvent (accessible via vapor pressure measurements) and the chemical potential of the solute (via the saturation concentration of the polymer in equilibrium with the pure crystalline state). Furthermore such systems offer an opportunity to learn more about the occurrence of non-equilibrium phenomena in mixtures containing crystallizable components. The freezing in of intermediate states during the

---

[a)]   Previously Ural State University, Institute of Physics and Applied Mathematics,
       Lenin Ave.51, Ekaterinburg, 620083 RUSSIA

[b)]   Author to whom correspondence should be addressed.
       Electronic mail: Bernhard.Wolf@Uni-Mainz.de

approach of equilibria is well documented for polymer blends of semi-crystalline and amorphous components in general[1-13] and with PEO as one component in particular[14-19]. Analogous investigations for the considerably less viscous solutions of such polymers (which should therefore be less prone to non-equilibria phenomena) are to our knowledge lacking.

For the reasons outlined above we have studied the system $CHCl_3$/PEO (PEO: polyethylene oxide) and prepared mixtures of constant composition in several different manners. The vapor pressures of the solutions that are in the individual cases established upon standing and the degrees of crystallinity of the coexisting pure polymer yield answers to the question, whether the particular mode of preparation of the solution affects its behavior. Calorimetric studies[20], vapor sorption experiments[21-23] and inverse gas chromatography measurements[24] performed for this system were very helpful for this purpose.

## II. EXPERIMENTAL PROCEDURES AND THEORETICAL BACKGROUND

### A Materials

Poly(ethylene oxide) (PEO 1000) was purchased from Polysciences Inc., Germany. The weight average molecular weight of the polymer $M_w = 1000$ kg mol$^{-1}$ according to the producer. GPC measurement (universal calibration) in dimethylformamide yielded a value of 1170 kg/mol and a molecular non-uniformity U= ($M_w/M_n$)-1 of 1.6.

Chloroform (p.a., Riedel de Ha¸n, Seelze, Germany) was dried over 3Å molecular sieves. Toluene, dichloromethane (technical grade, Merck, Darmstadt, Germany) were freshly distilled before use.

### B Film preparation

All mixtures of PEO and solvent studied thermodynamically were prepared from thin films of the polymer. To obtain these foils, homogeneous solutions containing 1 wt% PEO in chloroform, dichloromethane or toluene were cast on Teflon. The solvent was slowly evaporated at a room temperature and the remaining polymer was then dried to constant weight at 25°C under vacuum. The amount of the casting solution was chosen such that approximately 50-60 μm thick films are obtained. This dimension was chosen because one knows from literature[25] that it is still high enough to guarantee bulk properties of the polymer. In addition to solvent quality we have also checked whether the thermal history of the polymer sample plays a role. To that end films obtained from chloroform were molten, kept at 100 °C for several hours and then dipped in liquid nitrogen until they were totally frozen. The resulting films were visually clear and kept in the refrigerator until use.

### C DSC

The enthalpies of melting were determined by means of differential scanning calorimeter (DSC-7, Perkin-Elmer) with a heating rate of 10 °C min$^{-1}$. For these measurements we have used special pans that can be sealed, in order to avoid loss of solvent. By weighting in the required amounts of small pieces of the PEO films plus solvent, the mixtures of different compositions were prepared at room temperature. The samples were then equilibrated for at least one week. Orienting experiments have demonstrated that the DSC results after 1 week of standing are identical with results obtained after 2 weeks. Before and after the DSC experiments the pans were weighted to control that no solvent was lost.

### D HSGC

The combination of headspace sampling and gas chromatography for the determination of vapor pressures is already described in detail[26]. The measurements were performed at 25 °C using a pneumatically driven thermostatted head-space sampler (Dani HSS 3950, Milano, Italy) which takes 50 μL of the equilibrium gas phase and injects this mixture of solvent and air into a gas chromatograph (Shimadzu GC 14B, Kyoto, Japan). The amount of solvent contained in the sample volume – being proportional to the vapor pressure – is detected by a thermal conductivity detector and registered by means of an integrator (Shimadzu, Chromatopac C-R6A). We have tested the necessity to correct for the imperfection of the gas[27] (taking the required substance specific parameters from the literature[28]) and found that such influences can be neglected. For $CHCl_3$ we kept the column at 80 °C and that of the injector plus detector at 120 °C. The capillary column AT-WAX (Alltech Associates Inc., Deerfield, USA) had a length of 15 m, a diameter of 0,53 mm and a film thickness of 2,5 μm. In order to obtain reliable equilibrium data we have applied the method of Multiple Head-Space Extraction (MHSE)[29,30].

Larger tubes (now 20 cm³) were used for the present purposes. Approximately 0.25 g of the polymer films obtained from different solvents or after special thermal treatment were coiled up together with an aluminum foil, punching little holes though support and polymer film to increase the surface. These coils where than inserted into special glass tubes and exposed to an atmosphere saturated with the vapor of $CHCl_3$ until the polymer has taken up the required amount. The tubes were then sealed with septa for the vapor pressure measurements and equilibrated for one more week on a roll-mixer. After completion of the measurement further solvent was added via the gas phase and the procedure repeated until the difference between the vapor pressure above the solution and the vapor pressure of the pure solvent becomes too small to yield reliable data.

### E  Optical microscopy

The morphologies of the different PEO-films (sample thickness 50-60 μm) were observed at room temperature using polarized light by means of the microscope Olympus BX50, equipped with a charge coupled device (CCD) camera for digitalization of the images. In order to obtain information on the PEO particles that coexist with solutions of moderate polymer contents, we have repeated some HSGC experiments in the absence of the aluminum foil in small vessels so that we can place the mixture under the microscope.

## III. RESULTS

### A  DSC

The degrees of crystallinity, $a$, of the polymer contained in the different mixtures were calculated from $\Delta H_m$, the heats of melting measured per gram of PEO according to the following equation

$$a = \frac{\Delta H_m}{\Delta H_m^o} \qquad (1)$$

in which $\Delta H_m^o$ is the heat of melting per gram of 100% crystalline PEO, for which the literature reports the values of 205 J/g[31] and 203 J/g[32]. The parameter $a$ can also be written as

$$a = \frac{c}{c+a} \qquad (2)$$

where $c$ and $a$ represent the masses of crystalline and amorphous polymer, respectively.

Fig. 1 gives the results for experiments that differ only in the production of the PEO films used for the measurements. This graph demonstrates that the details of sample preparation play an important role on the degrees of crystallinity observed via DSC at high polymer concentrations. The values are largest if the films are prepared from solutions in toluene and smallest if the polymer melt is quenched by means of liquid nitrogen.

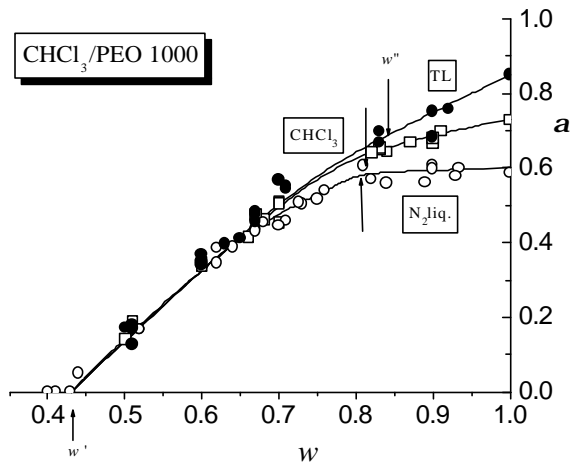

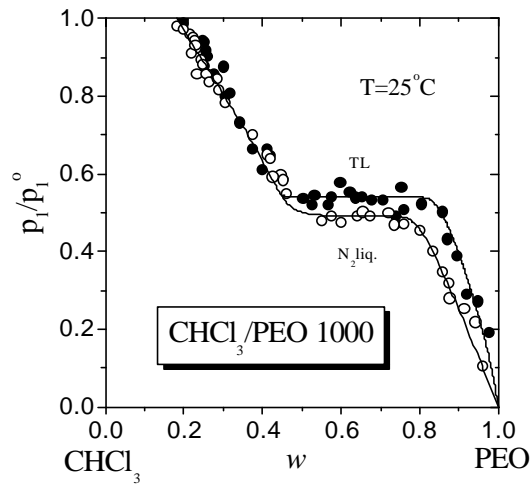

Fig. 1: Dependence of the degrees of crystallinity, $\alpha$ on the weight fraction, $w$, of PEO in mixtures with $CHCl_3$; how the polymer films used for these measurements have been prepared is indicated at the different curves. TL and $CHCl_3$: cast from these solvents; $N_2$ liq: films prepared from $CHCl_3$ where heated to 100 °C for several hours, subsequently quenched by means of liquid nitrogen and kept at low temperature up to the measurement. At the composition $w''$ (indicated by arrows) the vapor pressure of $CHCl_3$ becomes constant and at $w'$ the last solid PEO particles disappear.

Fig. 2: Reduced vapor pressures of $CHCl_3$ as a function of the weight fraction of PEO for films prepared from solutions in toluene and for films quenched from the melt by means of liquid nitrogen.

## B  HSGC

The different modes of sample preparation also affect the vapor pressures $p_1$ of $CHCl_3$ as demonstrated very clearly by the following graphs. In all cases the composition dependence of the reduced vapor pressures (normalized to $p_1^o$, the vapor pressure of the pure solvent) can be subdivided into three part: A steep ascend resulting from the addition of solvent on the polymer rich side is separated from a less pronounced increase on the solvent side by a region within which $p_1$ does not depend on composition. Below $w \approx 0.2$ the difference between the vapor pressure above the solution and that of the pure solvent lies within experimental error.

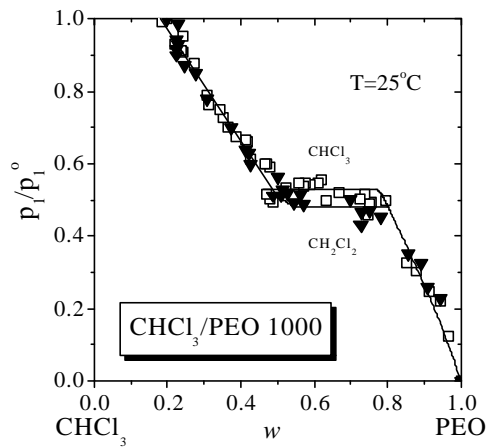

Fig. 3: As Fig. 2 but for films prepared from $CHCl_3$ or from $CH_2Cl_2$, respectively.

## IV. DISCUSSION

In this section we want to address several questions the experimental findings have raised. First of all it would be desirable to understand why the particular method of film making does not only influence the degrees of crystallinity of the pure PEO but also that prevailing in mixtures of low solvent content. Another question, that is closely related to the previous one, refers to the composition dependence of the vapor pressure of $CHCl_3$. Here we are interested to find out whether the measured vapor pressures monitor the same fraction of pure PEO (not taken up by the solvent) as the DSC experiments. Finally we want to check the consistency of thermodynamic information by comparing the directly observable saturation concentration of PEO in $CHCl_3$ with that predicted by the interaction parameters resulting from the vapor pressure measurements.

### A  Effects of sample preparation

The composition dependence of *a* shown in Fig. 1 reveals that the PEO films prepared from toluene contain considerably more crystalline material than that resulting from quenching the polymer melt. The data for films cast from $CHCl_3$ solution fall between. These observations can be rationalized in at least two manners: In terms of solvent quality and by means of kinetic considerations. According to the first option the tendency to form crystals should increase as the thermodynamic quality of the solvent falls. This reasoning is in agreement with the fact that toluene is much more unfavorable[24] than $CHCl_3$. The lower volatility of TL, on the other hand, would provide more time for crystallization. These findings are in accord with reports[33,34] on the crystallinity of PEO in blends with amorphous polymers prepared from different solvents. The lowest *a* values observed for the quenched sample reflect the large extent of supercooling and insufficient time for crystal growth.

Less clear cut is the explanation of the gradual reduction of the degree of crystallization observed with the solvent cast films upon the addition of $CHCl_3$ in the region 1 < *w* < *w*" (compare Fig. 4), i.e. before the vapor pressure become independent of composition. Intuitively one would have expected *a* to be constant, like in the case of the quenched film, because of the notion that solvent should under these conditions be taken up exclusively by the amorphous zones of PEO, leaving the crystallites untouched. On the basis of the observation that *a* approaches identical values as *w* falls to *w*" one may speculate that the larger *a* values for the solvent cast films represent a situation, which is still farther from equilibrium than that realized in quenching experiments. In view of the fact that poor solvents increase the crystallization tendency by dispelling the polymer chains out of the dissolved state, this interpretation appears thinkable, despite the fact that equilibrium considerations are doubtful and may apply locally only.

### B  Vapor pressure and state of the polymer

Fig. 4 sketches the typical variation of the vapor pressures with composition and the corresponding appearance of the mixtures under the microscope. The most obvious feature, already reported before[21], is the constancy of $p_1$ within a certain composition range of the mixture. It implies that the chemical potential of the solvent (and consequently also of the solute) does not change within this interval and thus indicates the existence of a phase equilibrium within the composition range between *w*' and *w*". This permanence of vapor pressure indicates the coexistence of PEO crystals with the saturated polymer solution. The increase in vapor pressure observed in the region between *w* = 1 and *w*" is ascribed to the only partially crystalline nature of the PEO. Small amounts of solvent added to the pure polymer are at first exclusively incorporated into its amorphous parts. What we measure under these conditions is therefore the vapor pressure of the mixed phase created by the non-crystalline parts of the polymer and solvent. This situation prevails until *w* reaches *w*", the saturation concentration of crystalline PEO in the liquid mixture. Only after *w* falls below *w*" the polymer crystals start to dissolve upon the addition of further solvent. As the composition of the total system moves from $w''$ to $w'$ one can observe under the microscope that the number of crystallites decreases and that the surviving ones become considerably larger, probably because the solvent supports the reconstruction of crystals leading to larger and more perfect structures.

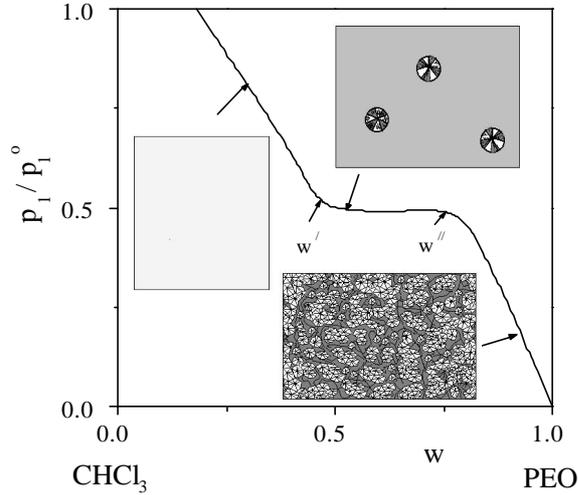

Fig. 4: Typical dependence of the reduced vapor pressure of CHCl$_3$ on the weight fraction of PEO and of the morphologies observed within the different composition ranges. Starting from pure PEO $p_1/p_1^o$ rises steadily as solvent is added until $w''$ is reached; within this area one observes a dense arrangement of comparatively small spherulites. As w under-runs $w''$, the pressure remains constant and individual crystalline particles, which grow as w falls, become visible. For $w < w'$, the mixture is homogeneous and $p_1/p_1^o$ rises steadily again.

In view of the above-said it is obvious that the particular manner, in which the vapor pressure increases as the concentration of PEO in the mixtures falls, should yield information on the fraction of the polymer that is under certain condition crystalline. For this reason the measured vapor pressures ought to offer additional information on the degree of crystallinity. The following considerations demonstrate, how the data may be evaluated in this respect.

For $w > w''$ only the amorphous part of the polymer can take up solvent and form a mixed phase. It is exclusively this portion of the system, which participates in liquid/vapor equilibria. In other words: The measured vapor pressure yields information on the (local) polymer concentration in the mixed phase, termed $e$. The vapor pressure $p_1$ is therefore much higher than expected from the over-all concentration $w$. Introducing **b** as the fraction of the polymer that does not become part of the solution under given conditions (by analogy to the degree of crystallinity **a** of Eq. (2)) we can express the local concentration $e$ in terms of the overall concentration $w$ and the fraction $(1 - \boldsymbol{b})$ of the polymer that is amorphous as

$$e = \frac{w(1-\boldsymbol{b})}{w(1-\boldsymbol{b})+(1-w)} \quad (3)$$

Rearrangement of this relation yields the following expression for the part of the polymer that does not participate in the liquid/vapor equilibrium

$$\boldsymbol{b} = \frac{w-e}{w(1-e)} \quad (4)$$

Inside the interval $w' < w < w''$, within which the vapor pressure does not depend on composition, $e$ is identical with $w'$ and it is easy to calculate **b**; for higher polymer contents the equilibrium vapor pressure curve needs to be extrapolated. An analysis of all results demonstrates that the best representation of the reduced vapor pressures as a function of $w$ within the homogeneous range ($w < w'$) can be achieved by the following simple relation

$$\frac{p_1}{p_1^o} = A(1-w)^B \quad (5)$$

which automatically fulfills the requirement of vanishing vapor pressure in the limit of the pure polymer; for the present system the parameters read $A = 1.32$ and $B = 1.27$.

By means of Eq. (5) it is possible to extrapolate the $p_1/p_1^o$ values measured for homogeneous solutions into the experimentally inaccessible part ($w > w''$), where a liquid phase coexists with polymer crystals. Fig. 5 shows an example for this procedure, namely for films prepared by quenching PEO melts with liquid nitrogen. From the curve identified by the letter $e$ it can be read what the vapor pressures of highly concentrated solutions would be in the (hypothetical) absence of crystal formation, i.e. complete dissolution of the polymer.

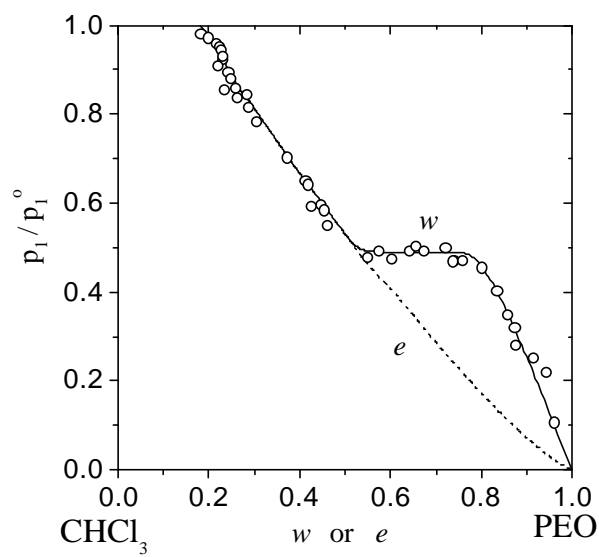

Fig. 5: Extrapolation of the vapor pressures measured for homogenous mixtures (data for PEO quenched from the melt by means of liquid nitrogen) into the region of high polymer concentration, where the system consists of pure PEO and a solution of PEO in $CHCl_3$, by means of Eq. (5).

Reading the corresponding pairs of $e$ and $w$ (constant reduced vapor pressures) from graphs of the above type within the interval $w' < w < 1$, we can by means of Eq. (4) calculate $b$, the fraction of PEO that does not become part of the solution under given conditions. For simplicity we call $b$ henceforth "degree of insolubility", by analogy to $a$, the degree of crystallinity. Fig. 6 demonstrates how the portion of polymer that is unable to mix with the solvent becomes less as $w$ decreases. Only below $w'$ all PEO present in the system is dissolved. For the assessment of the results one should keep in mind that the parts of the curves represented by full symbols are not based on an extrapolation according to Eq. (5); they are directly accessible from experimental data.

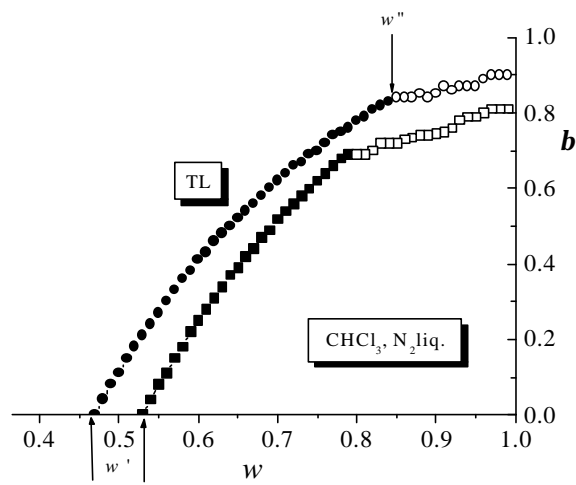

Fig. 6: Dependence of β, the degree of insolubility, on the weight fraction of PEO calculated by means of Eq. (4) for the system $CHCl_3$/PEO. The different modes of film preparation are indicated in the graph and described in the text. Full symbols: directly obtained from measured vapor pressures, open symbols: calculated by means of Eq. (5).

## C  Comparison of HSGC and DSC information

According to first considerations $b$ should be identical with $a$ because of the fact that the crystalline polymer is unable to take up solvent. Only the part that is required to reach the (normally comparatively low) saturation concentration is transferred into the solution, the rest remains in the pure, crystalline state. The following three graphs demonstrate that this simple argumentation does not hold true and that $b$ is under almost all conditions larger than $a$.

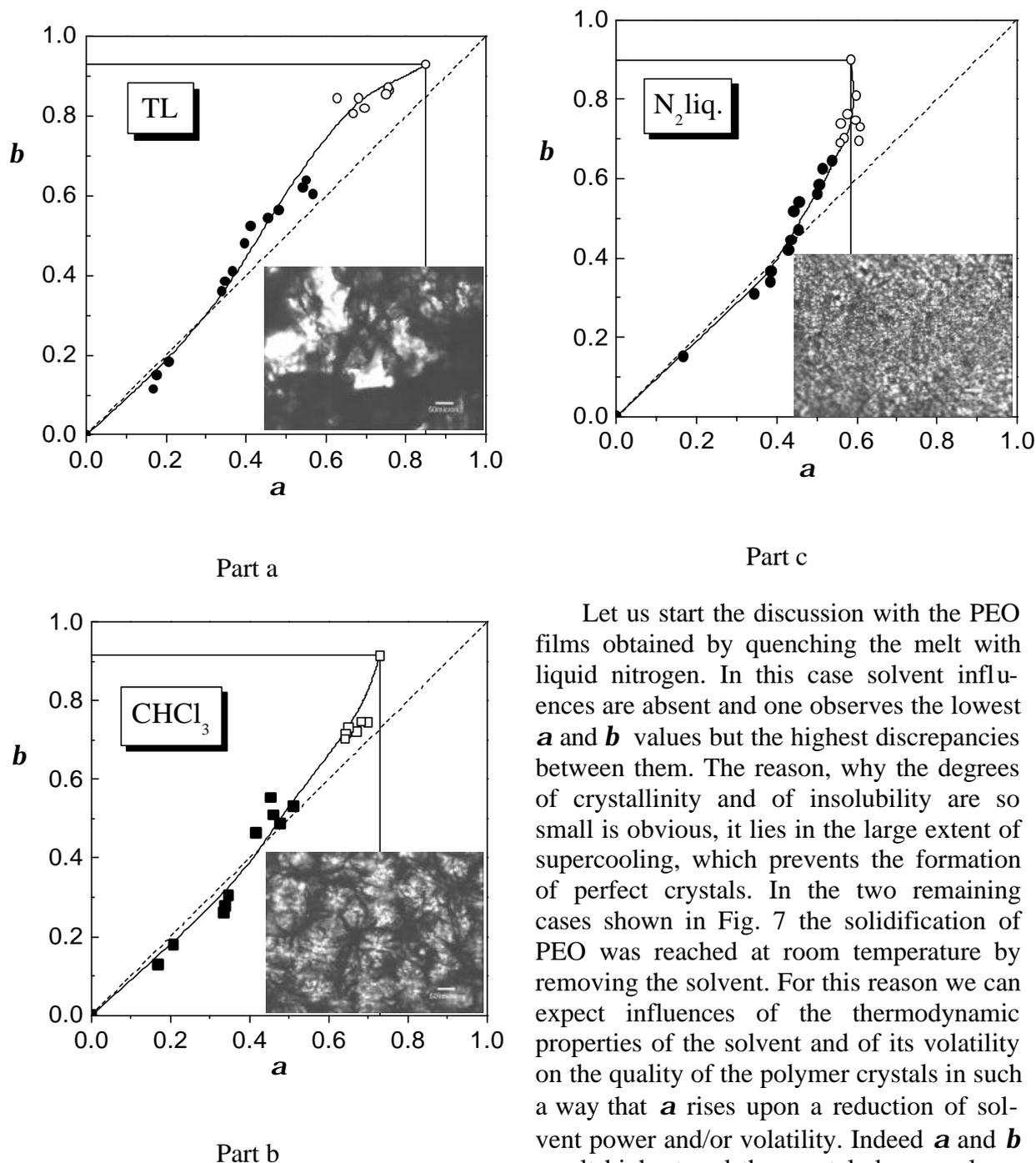

Part a

Part b

Part c

Fig. 7: Interdependence of β, the degree of insolubility, and *a*, the degree of crystallinity, for the system CHCl$_3$/PEO and the different modes of preparing PEO films. Full symbols: *b* directly obtained from measured vapor pressures, open symbols: *b* calculated by means of Eq. (5). The inserts show the morphologies of the particular PEO films used for these measurements. *Part a*: cast from solutions in toluene, *part b*: cast from solutions in CHCl$_3$, and *part c*: quenched from the polymer melt.

Let us start the discussion with the PEO films obtained by quenching the melt with liquid nitrogen. In this case solvent influences are absent and one observes the lowest *a* and *b* values but the highest discrepancies between them. The reason, why the degrees of crystallinity and of insolubility are so small is obvious, it lies in the large extent of supercooling, which prevents the formation of perfect crystals. In the two remaining cases shown in Fig. 7 the solidification of PEO was reached at room temperature by removing the solvent. For this reason we can expect influences of the thermodynamic properties of the solvent and of its volatility on the quality of the polymer crystals in such a way that *a* rises upon a reduction of solvent power and/or volatility. Indeed *a* and *b* result highest and the crystals become largest for film preparation from toluene, which is the worst of the present solvents[24] and the least volatile.

One essential finding still requires discussion, namely the observed discrepancies between the degrees of crystallinity (DSC) and the degrees of insolubility (vapor pressures). The most probable explanation is offered by the assumption that part of the amorphous PEO, not registered in the caloric experiments, is trapped inside of crystalline material and thus inaccessible to the solvent. This would explain, why *b* is always considerably larger than *a* at very high polymer concentrations and why these differences decrease with rising solvent content of the mixture. The dissimilarities in *b* and *a* that survive as *w* approaches *w*' lie probably within experimental error. In order to check the just formulated hypothesis, we have looked at the particles of solid PEO more precisely and have indeed observed more or less pronounced dark regions in their center, which are tentatively interpreted as trapped amorphous regions. Examples for such micrographs are shown in Fig. 8. Further experiments are required to confirm the occlusion of non-crystalline material.

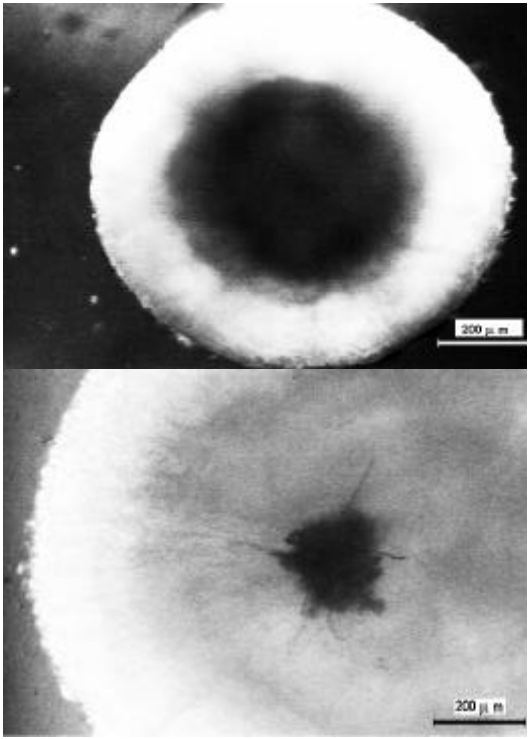

Fig. 8: Typical appearance of crystallites within the composition region of constant vapor pressure for the system $CHCl_3$/PEO (crossed polarizers). The example refers to film preparation from solutions in $CHCl_3$.

## D   Interaction parameters and polymer solubility

The vapor pressures of $CHCl_3$ measured for polymer concentrations below *w*' are within experimental error independent of the type of sample production. For this reason it appears worthwhile to check whether it is possible to predict *w*', the saturation concentration of PEO, from the Flory-Huggins interaction parameter *c*, obtained from the measured vapor pressures of the solvent. This evaluation is performed by means of the relation

$$\ln \frac{p_1}{p_1^0} = \ln a_1 =$$
$$= \ln(1 - \varphi_P) + \left(1 - \frac{1}{N_P}\right)\varphi_P + c\varphi_P^2 \quad (6)$$

in which $a_1$ stands for the activity of the solvent (with the present systems there is no need to correct for the imperfections of the equilibrium vapor pressure) and $\varphi_P$ is the volume fraction of the polymer. Due to the similar densities of solvent and solute the differences between volume fractions and weight fractions can be neglected in view of experimental uncertainties. $N_P$, the number of segments of the polymer, was calculated from the molar volumes, defining that of solvent as the volume of a segment.

According to the original ideas of Flory and Huggins, *c* should be independent of composition, which is in reality rather the exception than the rule. How the differential interaction parameter describing the chemical potential of the solvent varies with composition in the present case is shown in Fig. 9

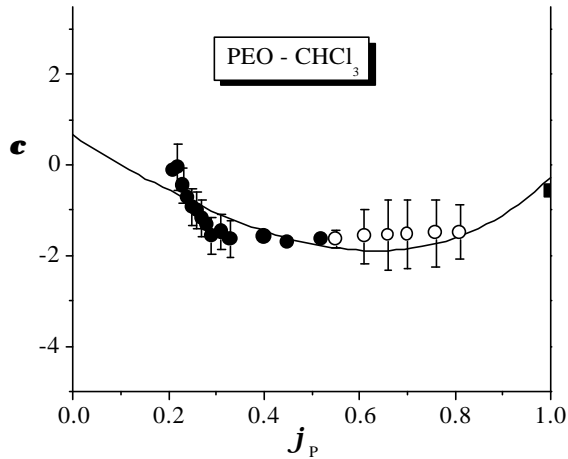

Fig. 9: Composition dependence of the Flory-Huggins interaction parameter $c$ for the system $CHCl_3$/PEO at 25 °C as obtained from the vapor pressure data shown in Fig. 2 and Fig. 3. The full circles correspond to homogenous mixtures, the open ones are based on the extrapolation of these values into the two-phase regime by means of Eq. (5). The full square represents the limiting value of $c$ for vanishing solvent content; it stems from measurements by means of inverse gas chromatography reported in the literature[24] for 100 °C. The curve is calculated by means of Eq. (7) with $z = 11.9$ and $n = 0.387$.

The individual data points of Fig. 9 are very different in their accuracy, as can be read from the error bars shown in this graph. The information stemming from the vapor pressures of homogeneous solutions (full symbols) is very precise for high polymer concentration, but the errors increase markedly as $w$ falls. Despite the fact that we do not have experimental data for $c_o$, it is evident from phenomenological considerations that the limiting value of $c$ for infinite dilution must be on the order of 0.5, because of the fact that the Flory-Huggins equation has to model colligative properties. The data for high $w$ values are all rather uncertain due to the necessity of an extrapolation procedure (Eq. (5)). Also incorporated in Fig. 9 is an interaction parameter based on inverse gas chromatography, i.e. the limiting value of $c$ for $w \to 1$ as measured at 100 °C where the polymer is liquid. It is obvious that data for different temperatures above the melting point of PEO should have been extrapolated to 25 °C to complement the present dependence. Unfortunately this information is, however, not available. For the analytical representation of $c(j)$ this data point was therefore discounted. Nevertheless it is obvious that the composition dependence of the Flory-Huggins interaction parameter must pass a minimum.

Such a behavior, which is by no means uncommon[26], cannot be rationalized by well established theories. Recently is could be explained in terms of chain connectivity and conformational variability of polymers[35,36]. This approach leads in its simplified form to the following uncomplicated expression

$$c \approx \frac{a^*}{(1-nj)^2} - zl(1+2j) \qquad (7)$$

in which $l = 0.5$ for truly high molecular weight polymers. Furthermore $a^*$ and $z$ are not independent of each other, but interrelated by the following expression, which holds in very good approximation true for most vinyl polymers[35,36]

$$a^* = 0.513 z + 0.5 \qquad (8)$$

A test of the versatility of Eq. (7) has demonstrated that this simple relation is capable of modeling a large diversity of fundamentally different composition dependencies[35,36], including the occurrence of minima, by the adjustment of two parameters only (cf. Fig. 9).

Interaction parameters, quantifying the effects of contact formation between polymer segments and solvent molecules, cannot only be obtained from equilibria between the solution and the pure vapor, but also from equilibria between the solution and the pure, crystalline polymer. The activity of the polymer in the solution can be formulated in terms of the original Flory-Huggins theory as

$$\ln a_P = \ln j_P + (1 - N_P)(1 - j_P) + x N_P (1 - j_P)^2 \quad (9)$$

by analogy to Eq. (6). The symbol $x$ instead of $c$ is used because of the fact that the numerical values of $c$ and $x$ disagree for a given polymer solution in case of composition dependent interaction parameters. According to the laws of phenomenological thermodynamics the integration of the interaction parameter $c$ yields the integral interaction parameter $g$ from which $x$ is obtained upon differentiation with respect to the polymer[37]. According to this procedure Eq. (7) yields the following expression for the concentration dependence of $x$

$$x \approx \frac{a}{(1-n)(1-nj)^2} - 2zl(1+j) \quad (10)$$

Under equilibrium conditions the chemical potential of those components, which are able to cross the phase boundary, must become identical in the coexisting phases. In case of liquid/vapor equilibria this requirement has lead to Eq. (6). For the liquid/solid equilibria of present interest it yields[38]

$$\frac{\Delta H_m}{RT_m} - \frac{\Delta H_m}{RT} = \ln j_P + (1 - N_P) j_P + N_P x (1 - j_P)^2 \quad (11)$$

$\Delta H_m$ is the segment molar heat of melting for 100 % crystalline PEO (8.4 kJ per mol of segments[39]) and $T_m$ is the melting temperature of the pure polymer (353.3 K). By means of Eqs. (10) and (11) it is now possible to determine the saturation concentration of the polymer, $j_{P,sat} \approx w'$, from the equality of the actual $x$ value with that required for the establishment of an equilibrium between the dissolved polymer and its pure crystals. This quantity can be obtained most easily by plotting the actual function $x(j)$ according to Eq. (10), containing the system specific information obtained from the vapor pressure measurements, in the same diagram as the $x$ values required to fulfil equilibrium condition of Eq. (11). The $j_{P,sat}$ value predicted in this manner by the vapor pressure measurements can be read from the intersect of these two curves (cf. Fig. 10).

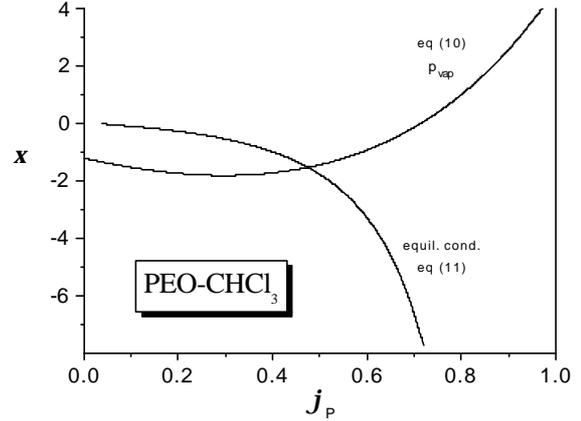

Fig. 10: Graphical determination of the saturation concentration of PEO in $CHCl_3$ (in equilibrium with PEO crystals) from the equality of the measured (differential) interaction parameter $x$ (Eq. (10)) with that required for equilibrium conditions (Eq. (11)).

The above evaluation leads to $j_{P,sat} = 0.48$, as compared with the directly observed value, which lies within the interval from 0.47 to 0.53, as can be read from Fig. 2 and Fig. 3. In view of the numerous unavoidable experimental uncertainties the agreement is remarkably good.

## V. CONCLUSIONS

According to the present results it is possible to distinguish three clearly separable composition ranges. Upon the addition of $CHCl_3$ to solid PEO the vapor pressure increases steadily within range *I* ($1 > w > w''$) up to a characteristic limiting value located well below that of the pure solvent. Within range *II* ($w'' > w > w'$) $p_1$ remains constant, despite the addition of further solvent. Finally, within range *III* (*w'* > *w* > 0) the vapor pressure rises again and approaches the value of the pure solvent. Range *I* should be absent for fully crystalline polymers, because ideal crystals do not swell and the first seizable vapor pressure for a solution that is in equilibrium with the pure polymer must be that of the saturated solution (*w'*). The existence of range *I* is due to the amorphous parts of PEO, which can take up solvent until *w'* is reached. Range *II* results from the coexistence of the saturated solution with variable amounts of polymer crystals. Finally, no solid material is available in range *III* and we are back to the normal situation encountered with the solutions of amorphous polymers. According to the present results it is difficult to reach thermodynamic equilibria within range *I*. Vapor pressures, degrees of crystallinity *a* and degrees of insolubility *b* depend markedly on the details of sample preparation. No problems with the attainment of equilibria are observed within range *III*. Range *II* assumes an intermediate position in this respect.

The observed non-equilibrium behavior, i.e. the influences of the history of the PEO films on *a* (as documented by micrographs) and on the variation of *a* and *b* with composition are interpreted in terms of local and temporal equilibria, which are frozen in during the removal of solvent or during the quenching process. The general finding *b* > *a* and the fact that these differences decrease upon dilution are tentatively interpreted as a trapping of amorphous PEO inside the crystalline material during sample preparation and its gradual release by the addition of solvent. This hypothesis is supported by micrographs pointing at the existence of such occlusions.

Equilibrium information was obtained from two sources: From the vapor pressures of the solvent above the solutions within zone *I* (chemical potential of the solvent) and from the saturation composition *w'* of the polymer (chemical potential of the polymer). The thermodynamic consistency of these data could be documented by predicting *w'* (liquid/solid equilibrium) from the liquid/gas equilibrium. These calculations have been made possible by a new approach, which allows a realistic modeling of the composition dependence of interaction parameters by means of only two adjustable parameters.

## Acknowledgements

The authors are grateful to the "Deutscher Akademischer Austauschdienst" and to the Civilian Research and Development Foundation (USA, REC-005.2001) for financial support.